\documentclass[a4paper,aps,floats,intlimits,pre,notitlepage,intlimits,superscriptaddress,8pt,floatfix,twopages,twocolumn]{revtex4-1}
\usepackage{amsmath,amsthm,amssymb,amsthm,dsfont,mathrsfs,bbm}
\usepackage{mathtools,multirow,xfrac}
\usepackage[caption=false]{subfig}
\usepackage[english]{babel}
\usepackage{hyperref}
\hypersetup{pdfauthor={Malatesta Parisi Sicuro},pdftitle={Large deviations for matching},%
            colorlinks, linktocpage=true, pdfstartpage=3, pdfstartview=FitV,%
    breaklinks=true, pdfpagemode=UseNone, pageanchor=true, pdfpagemode=UseOutlines,%
    plainpages=false, bookmarksnumbered, bookmarksopen=true, bookmarksopenlevel=1,%
    hypertexnames=true, pdfhighlight=/O,%
    urlcolor=orange, linkcolor=blue, citecolor=red, 
        }

\usepackage{booktabs}
\usepackage{braket}
\usepackage{pgfplots}

\usepackage{lmodern}
\usepackage[cal=esstix]{mathalfa}
\usepackage[T1]{fontenc}

\newcommand{\R}{\mathds{R}}

\newcommand{\N}{\mathds{N}}

\newcommand{\dd}{\mathrm{d}}
\newcommand{\e}{\mathrm{e}}

\newcommand{\mediaE}[1]{{\mathbb E\left[#1\right]}}

\catcode`,\active

\catcode`\,12

\theoremstyle{plain}

\usepackage{tikz}
\usetikzlibrary{patterns,decorations,arrows,shapes.geometric,arrows,mindmap,backgrounds,positioning,automata,positioning,fit,backgrounds,positioning,arrows,matrix,calc}

\tikzset{
  common/.style={draw,name=#1,node contents={},inner sep=0,minimum size=2},
  disc/.style={circle,common=#1},
  square/.style={rectangle,common={#1}},
}

\allowdisplaybreaks
\begin{document}
\title{Fluctuations in the random-link matching problem}
\author{Enrico M.~Malatesta}\affiliation{Bocconi Institute for Data Science and Analytics, Bocconi University, Milano 20136, Italy}
\author{Giorgio Parisi}\affiliation{Dipartimento di Fisica, INFN -- Sezione di Roma1, CNR-IPCF UOS Roma Kerberos, Sapienza Universit\`a di Roma, P.le A. Moro 2, I-00185, Rome, Italy}
\author{Gabriele Sicuro}\email{gabriele.sicuro@for.unipi.it}\affiliation{Dipartimento di Fisica, Sapienza Universit\`a di Roma, P.le A. Moro 2, I-00185, Rome, Italy}
\begin{abstract}Using the replica approach and the cavity method, we study the fluctuations of the optimal cost in the random-link matching problem. By means of replica arguments, we derive the exact expression of its variance. Moreover, we study the large deviation function, deriving its expression in two different ways, namely using both the replica method and the cavity method. \end{abstract}
\maketitle
\section{Introduction} The application of replica theory \cite{mezard1987spin} to the study of random combinatorial optimization problems (RCOPs) has a long tradition that started more than thirty years ago with the seminal works by \textcite{Orland1985} and \textcite{Mezard1985}. It became immediately clear that methods borrowed from the theory of disordered systems are very effective to study the average properties of ensembles of RCOPs. Exact results have been obtained, e.g., for mean-field versions of many RCOPs, such as the matching problem \cite{Mezard1985,*Mezard1988,*Lucibello2017,*Lucibello2018}, the travelling salesman problem \cite{Mezard1986,*Krauth1989a}, $K$-SAT problems \cite{Mezard2002,*Mezard2002b,*Braunstein2007,*Krzakala2007}, graph partitioning \cite{Fu1986,*Kanter1987}, matching enumeration in sparse graphs \cite{Zdeborova2006}, constraint least square problems \cite{Fyodorov2014}, and many others. In the case of the matching problem, expressions for the finite-size corrections to the average optimal cost have also been obtained \cite{Mezard1987,*Parisi2002,*Caracciolo2017}. The parallel success of the cavity method \cite{mezard1987spin,Mezard2001,*Mezard2003} inspired message-passing algorithms for the solution of specific instances of RCOPs \cite{mezard2009information}.

In this paper, we study the deviations from the average optimal cost (AOC) in a particular RCOP, the random-link matching problem, using both the replica and the cavity method. Such an investigation is of methodological interest beyond the analysis of the specific problem. Replica calculations for the study of large deviations of thermodynamic-like functionals in presence of disorder are quite rare in the literature. Indeed, except for the analysis in Ref.~\cite{Fyodorov2014}, where the replica method is applied to the study of large deviations of the minimum, the results cited above mainly concern the typical properties of the solutions only. On the other hand, the application of the cavity method to the study of large fluctuations has not been explored up to now, with the exception of Ref.~\cite{Rivoire2005}. Here we apply, for the first time, both approaches to the same problem, showing that they lead to the same result.

We will study both the small fluctuations around the AOC and the large deviations from it. In the matching problem we assume that $2N$ vertices, labelled by the index $i\in\{1,\dots,2N\}$ are given, alongside with a positive weight $w_{ij}\in\R^+$ for each pair $(i,j)$. We search for the symmetric matrix $\mathbf M=(m_{ij})_{i,j}$ such that the cost
\begin{equation}
 C_N[\mathbf M]\coloneqq\sum_{i<j}m_{ij}w_{ij}
\end{equation}
is minimized. The minimization has to be performed on the set of matrices $\mathbf M$ such that
\begin{equation}
 m_{ji}=m_{ij}\in\{0,1\}\ \forall i,j,\qquad \sum_{i=1}^{2N} m_{ij}=1\quad \forall j.
\end{equation}
In the random-link version of the problem, the quantities $w_{ij}$ are supposed to be i.i.d.~random variables, with probability density function $\rho(w)$. Using the replica theory, in Ref.~\cite{Mezard1985} it has been proven that, if $\lim_{w\to 0}\rho(w)=1$, then
{color{blue}\begin{equation}
 \mathcal C\coloneqq\lim_{N\to+\infty}\mediaE{\min_\mathbf{M}C_N[\mathbf M]}=\frac{\zeta(2)}{2},
\end{equation}}
where we have denoted the average over all possible realizations by $\mediaE{\bullet}$, and $\zeta(z)$ is the Riemann zeta function. The calculation was performed introducing a partition function
\begin{equation}
 Z_w(\beta)\coloneqq\sum_{\{m_{ij}\}}\prod_{i=1}^{2N}\mathbb I\left(\sum_{j=1}^{2N} m_{ij}=1\right)\e^{-\beta N C_N[\mathbf M]}
\end{equation}
where the indicator function $\mathbb I(\bullet)$ is equal to one if its argument is true, zero otherwise. From the expression above, the replicated free-energy can be derived
\begin{equation}\label{Phinbdef}
 \Phi(n,\beta)\coloneqq -\lim_{\mathclap{N\to +\infty}}\frac{\ln \mediaE{Z_w^n(\beta)}}{\beta n N}.
\end{equation}
 The functional $\Phi(n,\beta)$ has been obtained, in the replica symmetric hypothesis, in Ref.~\cite{Mezard1985} and it is equal to
\begin{subequations}
\begin{multline}
-\beta n \Phi(n,\beta)=-\frac{\beta}{2}\sum_{p=1}^n\frac{\Gamma(n+1)}{\Gamma(p)\Gamma(n-p+1)}q_p^2\\
+2\ln\left[\iint\frac{dx\,d\eta}{2\pi}(-i\eta)^n\exp\left(i\eta x+\sum_{p=1}^\infty\frac{x^pq_p}{p!}\right)\right ]\label{Phinb}
\end{multline}
where the order parameters $q_p$ have to be specified using the saddle-point condition
\begin{equation}
 \frac{\partial \Phi(n,\beta)}{\partial q_p}=0\qquad p\in\N.\label{sp}
\end{equation}
\end{subequations}
The asymptotic AOC is then recovered as the value of $\Phi$ in the zero temperature limit, taking the number of replicas $n$ going to zero,
{
\begin{equation}
 \mathcal C=\lim_{\beta\to+\infty}\lim_{n\to 0}\Phi(n,\beta).
\end{equation}
Observe that we should, in principle, take $n\to 0$ first, and then $\beta\to+\infty$, but, as usual in replica calculations, we will assume that the order of two limits can be safely inverted. Here and in the following, we denote by $C_N\coloneqq\min_\mathbf{M}C_N[\mathbf M]$ the instance-dependent optimal cost, and by $\varrho_N(C)$ its distribution. We expect $\rho_N(C)$ to concentrate, for $N\to+\infty$, on the asymptotic AOC $\mathcal C$, $\lim_{N\to+\infty}\varrho_N(C)=\delta(C-\mathcal C)$. In particular, we will show that the fluctuations around the AOC are Gaussian. Moreover, we expect $C_N$ to satisfy a large deviation principle, i.e.,
\begin{equation}
    -\frac{\ln \rho_N(C)}{N}=L(C)+O\left(\frac{1}{N}\right),
\end{equation}
where $L(C)$ is a strictly convex, positive function having $L(\mathcal C)=0$, known as the Cram\'er rate function. In some sense, $L(C)$ measures how rare is to find an optimal cost $C\neq \mathcal C$ for large values of $N$.}

Our computation of the large deviation function $L(C)$ for the random-link matching problem starts exactly from Eq.~\eqref{Phinb}. It is well known, indeed, that the replicated average free-energy $n\Phi(n,\beta)$ contains information not only on the average free energy, but also on its fluctuations \cite{Crisanti2014}. In particular, using Eq.~\eqref{Phinbdef} it is possible to show that, for finite $n$, $-n\beta\Phi(n,\beta)$ is the cumulant generating function for the free energy $\ln Z_w(\beta)$ \cite{Crisanti1992b}. This fact has been used, for example, by Parisi and Rizzo \cite{Parisi2008,*Parisi2009,*Parisi2010a}, to extract the large deviation function in the Sherrington-Kirkpatrick model. Thet confirmed the anomalous scaling of fluctuations of the free energy in the RSB phase predicted, near the critical temperature, by \textcite{Crisanti1992b}, on the basis of a previous result by \textcite{Kondor1983}. In the present paper, we are interested in the fluctuation of the ``ground state free-energy'' $C_N$ in the random-link matching problem, and therefore we have to take $\beta\to+\infty$.  The cumulant generating function of the optimal cost is then obtained as
\begin{multline}
 \alpha\Phi(\alpha)\coloneqq\lim_{\mathclap{\substack{\beta\to +\infty\\n\beta=\alpha}}}n\beta \Phi(n,\beta)=-\lim_{\mathclap{N\to+\infty}}\frac{\ln\mediaE{\e^{-\alpha N C_N}}}{N}\\=\lim_{N\to+\infty}\frac{1}{N}\sum_{k=1}^\infty (-1)^{k-1}\frac{\kappa_k\alpha^k}{k!},
\end{multline}
where $\kappa_k$ is the $k$-th cumulant of the random variable $NC_N=\lim_{\beta\to+\infty}\beta^{-1}\ln Z_w(\beta)$ \cite{Crisanti1992b}, the first order in $\alpha$ simply being the asymptotic AOC, $\alpha\Phi(\alpha)=\alpha\mathcal C+o(\alpha)$. If $\lim_{N}N^{-1}\kappa_2$ is finite and different from zero, i.e., $\kappa_2=2\sigma^2N$ with $\sigma^2=O(1)$, this implies that $\mediaE{(C_N-\mathcal C)^2}= \sigma^2 N^{-1}$, i.e., small fluctuations of the optimal cost are Gaussian. 

If $\alpha\Phi(\alpha)$ is differentiable and convex, by the G\"{a}rtner--Ellis theorem the Cram\'er function of $\varrho_N(C)$ is obtained as the Legendre-Fenchel transform of $\alpha\Phi(\alpha)$,
\begin{subequations}\label{legendret}
\begin{multline}
 L(C)\coloneqq -\lim_{\mathclap{N\to+\infty}}\frac{\ln\varrho_N(C)}{N}=\inf_{\alpha}\left[\alpha\Phi(\alpha)-\alpha C\right]\\
 =\alpha_C\Phi(\alpha_C)-\alpha_C C,\end{multline}
 with $\alpha_C$ such that
\begin{equation}
C=\left.\frac{\partial\left[\alpha\Phi(\alpha)\right]}{\partial\alpha}\right|_{\alpha=\alpha_C}.
\end{equation}\end{subequations}
{Vice-versa, the cumulant generating function can be obtained from the rate function using, once again, a Legendre-Fenchel transform,
\begin{subequations}\label{legendret2}
\begin{equation}
 \alpha\Phi(\alpha)=\inf_{C}\left[L(C)+\alpha C\right]
 =L(C_\alpha)+\alpha C_\alpha,\end{equation}
 with $C_\alpha$ such that
\begin{equation}
\alpha=\left.\frac{\partial L(C)}{\partial C}\right|_{C=C_\alpha}.
\end{equation}\end{subequations}
Intuitively, therefore, the quantity $\alpha^{-1}$ plays the role of a ``fluctuation scale'' with respect to the deviation $C-\mathcal C$. Indeed, assuming $L(C)$ strictly convex and nonnegative, such that $L(\mathcal C)=0$, then $\alpha=0$ corresponds to $C_\alpha=\mathcal C$. On the other hand, being $\partial_C L(C)<0$ for $C<\mathcal C$, it is clear that positive values of $\alpha$ corresponds to negative fluctuations of the cost, and negative values of $\alpha$ corresponds to positive deviations from the asymptotic AOC.
}
In the following, we will derive the exact value of $\sigma^2$, proving that the small fluctuation of the optimal cost are indeed Gaussian, and we will obtain an expression for the function $\Phi(\alpha)$ that we will solve numerically.

\section{Small fluctuations} \label{sec:small}
Let us start from the computation of the variance of the optimal cost. The small $\alpha$ expansion of $\alpha\Phi(\alpha)$ up to $o(\alpha^2)$ terms will provide us the first and the second cumulant of the optimal cost, i.e., its mean and its variance. Due to the fact that we are performing an expansion around $\alpha=0$, it is useful to recall the expression of the saddle-point value of the order parameter $q_p\equiv Q_p$ in this particular case, i.e., for $n\to 0$ and $\beta\to+\infty$. It has been shown in Ref.~\cite{Mezard1985} that, introducing the function
\begin{equation}\label{g0}
G_0(t)\coloneqq\sum_{p=1}^\infty\frac{(-1)^{p-1}\e^{\beta p t}}{p!}Q_p,
\end{equation}
the saddle-point condition for $n\to 0$ and $\beta\to+\infty$ reads
\begin{equation}
 G_0(x)=2\int_{-x}^\infty\e^{-G_0(t)}\dd t\Rightarrow G_0(x)=\ln\left(1+\e^{2x}\right).
\end{equation}
To obtain the expansion of $\Phi(\alpha)$, let us start from the double integral appearing in the argument of the logarithm in Eq.~\eqref{Phinb}, to be evaluated on the saddle-point $Q_p$ for small $\alpha$. We have
\begin{multline}
 \iint\frac{\dd x\,\dd\eta}{2\pi}(-i\eta)^n\e^{i\eta x+\sum_{p=1}^\infty\frac{x^pQ_p}{p!}}=\\
 =1+\sum_{k=1}^\infty\frac{\alpha^k}{\beta^k k!}\iint\frac{\dd x\,\dd\eta}{2\pi}\ln^k(-i\eta)\e^{i\eta x+\sum_{p=1}^\infty\frac{x^pQ_p}{p!}}.
\end{multline}
Using now the representation of the logarithm $$\ln(x)=\int_0^{+\infty}\frac{\e^{-t}-\e^{-xt}}{t}\dd t$$ we can write
\begin{multline}
 \lim_{\substack{\beta\to+\infty\\\beta n=\alpha}}\iint\frac{\dd x\,\dd\eta}{2\pi}(-i\eta)^n\e^{i\eta x+\sum_{p=1}^\infty\frac{x^pQ_p}{p!}}=\\
 =1+\alpha\int\left[\theta(-h)-\e^{-G_0\left(h\right)}\right]\dd h\\
 +\frac{\alpha^2}{2}\!\iint\!\dd h_1\dd h_2\!\left[\theta(-h_1)\theta(-h_2)-\theta(-h_1)\e^{-G_0(h_2)}\right.\\\left.-\theta(-h_2)\e^{-G_0(h_1)}+\e^{-G_0\left(\max\{h_1,h_2\}\right)}\right]+o(\alpha^2)\\
 =1+\frac{\zeta(2)}{4}\alpha^2+o(\alpha^2).
\end{multline}
Similarly, the first term in Eq~\eqref{Phinb} can be written as
\begin{multline}
\frac{\beta}{2} \sum_{p=1}^\infty\frac{\Gamma(n+1)}{\Gamma(p)\Gamma(n-p+1)}Q_p^2=\\
 =\frac{\alpha}{2}\sum_{p=1}^\infty(-1)^{p-1}\left[1-\frac{\alpha}{\beta}\mathrm H_{p-1}+o\left(\frac{\alpha}{\beta} \right )\right ]Q_p^2\\ 
 =\frac{\alpha}{2}\zeta(2)-\frac{\alpha^2}{2\beta}\sum_{p=1}^\infty(-1)^{p-1}\mathrm H_{p-1}Q_p^2+o(\alpha^2).
\end{multline}
where $\mathrm H_p$ is the $p$th harmonic number. If we now use the fact that $\mathrm H_{p-1}=\int_{0}^\infty\frac{\e^{-t}-\e^{-pt}}{1-\e^{-t}}\dd t$, then
\begin{multline}
 \lim_{\substack{\beta\to+\infty\\n\beta=\alpha}}\frac{\beta}{2} \sum_{p=1}^\infty\frac{\Gamma(n+1)}{\Gamma(p)\Gamma(n-p+1)}Q_p^2=\\=\alpha\frac{\zeta(2)}{2}+\alpha^2\int_{-\infty}^{+\infty}\dd h\e^{-G_0(h)}\int_0^{+\infty} G_0(h-t)\dd t+o(\alpha^2)\\=\alpha\frac{\zeta(2)}{2}+\alpha^2\frac{\zeta(3)}{2}+o(\alpha^2).
\end{multline}
Collecting all results we get
\begin{equation}
 \Phi(\alpha)=\frac{\zeta(2)}{2}-\frac{\zeta(2)-\zeta(3)}{2}\alpha+o(\alpha),
\end{equation}
implying that small fluctuations of the optimal cost in the random-link matching problem are Gaussian, with a variance given by
\begin{equation}\label{varianza}
\mediaE{(C_N-\mathcal C)^2}=\frac{\zeta(2)-\zeta(3)}{N}+o\left(\frac{1}{N}\right).
\end{equation}
Recently \textcite{Wastlund2010} has derived the variance of the random-link assignment problem with exponentially distributed random weights on a complete bipartite graph. His result, obtained using a purely probabilistic approach, coincides with Eq.~\eqref{varianza}, apart from a global factor $4$, due to the fact that the optimal cost of the assignment problem is twice the optimal cost of the matching problem for $N\to+\infty$. In our calculation, only the assumption $\lim_{w\to 0}\rho(w)=1$ for the weight probability density function has been used. 

\section{Large deviations via replicas} 
In the previous Section we have performed a small $\alpha$ expansion to extract the variance of the optimal cost $C_N$ for $N\gg 1$. To get instead the large deviation function $L(C)$, we have to keep $\alpha$ finite. We follow two different approaches and we compare then our results with numerical simulations. Let us start from the replica method. Following our general recipe, we derive the large deviation function starting from Eq.~\eqref{Phinb}, writing down the saddle-point equation in the $\beta\to+\infty$ limit, taking $n\beta=\alpha$ fixed. We start considering $\alpha<0$ (we will relax this assumption in our cavity calculation). Using the fact that, for $n\in(-1,0)$,
\begin{equation}
\int(-i\eta)^n\e^{i\eta x}\dd\eta=-\frac{2\Gamma(n+1)\sin(n\pi)}{(-x)^{n+1}}\theta(-x),
\end{equation}
{and defining as in Eq.~\eqref{g0}
\begin{equation}
    G_{n,\beta}(t)\coloneqq\sum_{p=1}^\infty\frac{(-1)^{p-1}Q_p\e^{\beta p t}}{p!}
\end{equation}
the argument of the logarithm can be re-written as
\begin{multline}
 \iint\frac{\dd x\,\dd\eta}{2\pi}(-i\eta)^n\exp\left(i\eta x+\sum_{p=1}^\infty\frac{x^p Q_p}{p!}\right)=\\=-\frac{\beta\Gamma(n+1)\sin(n\pi)}{\pi}\int_{-\infty}^{0}\exp\left(\sum_{p=1}^\infty\frac{x^p Q_p}{p!}\right)\frac{\dd x}{(-x)^{n+1}}\\
 =-\frac{\beta\Gamma(n+1)\sin(n\pi)}{\pi}\int_{-\infty}^{+\infty}\e^{-n\beta t-G_{n,\beta}(t)}\dd t.\end{multline}}
The saddle-point equation \eqref{sp} becomes
\begin{multline}\label{spnb}
\beta p\binom{n}{p}Q_p=2\frac{(-1)^p}{p!}\frac{\int_{-\infty}^{+\infty}\e^{(p-n)\beta t-G_{n,\beta}(t)}\dd t}{\int_{-\infty}^{+\infty}\e^{-n\beta t'-G_{n,\beta}(t')}\dd t'}\Longrightarrow\\
G_{n,\beta}(x)=-2\frac{\int\limits_{-\infty}^{+\infty} K_{n,\beta}(x+t)\e^{-n\beta t-G_{n,\beta}(t)}\dd t}{\int_{-\infty}^{+\infty}\e^{-n\beta t'-G_{n,\beta}(t')}\dd t'}.
\end{multline}
where the function
\begin{equation}
    K_{n,\beta}(u)\coloneqq \sum_{p=1}^\infty\frac{\e^{\beta up}}{\binom{n}{p}\beta p(p!)^2}
\end{equation}
appears. Observing now that (see Appendix)
\begin{equation}\label{limite}
 \lim_{\substack{\beta\to+\infty\\n\beta=\alpha}}K_{n,\beta}(u)=\theta(u)\left(\frac{1}{\alpha}-u\right),
\end{equation}
the saddle-point equation can be written as
\begin{subequations}\label{PhiG}
\begin{multline}\label{spG}
 G_\alpha(x)\coloneqq\lim_{\substack{\beta\to+\infty\\n\beta=\alpha}}G_{n,\beta}(x)\\
 =-\frac{2\alpha}{\mathcal Z_\alpha}\int_{-x}^{+\infty}\left(x+t-\frac{1}{\alpha}\right)\e^{-\alpha t-G_\alpha(t)}\dd t,
\end{multline}
where we have introduced
\begin{equation}
    \mathcal Z_\alpha \coloneqq -\alpha \int_{-\infty}^{+\infty}\e^{-\alpha t'-G_\alpha(t')}\dd t'.
\end{equation}
The equation above implies that $\lim_{x\to +\infty} x^{-1}G_\alpha(x)=2$ for any value of $\alpha$, i.e., $G_\alpha(x)\sim 2x$ for large $x$. Assuming $\alpha> -2$, this also implies that $\lim_{x\to-\infty}G_\alpha(x)=0$.  By consequence, the integral appearing in the expression of $\mathcal Z_\alpha$ converges for $\alpha> -2$ only and diverges otherwise. Using the saddle-point equation \eqref{spnb} in Eq.~\eqref{Phinb} we finally get
\begin{equation}\label{aPhi}
 \alpha\Phi(\alpha)=\frac{\alpha}{\mathcal Z_\alpha}\int_{-\infty}^{+\infty}G_\alpha(t)\e^{-\alpha t-G_\alpha(t)}\dd t-2\ln\mathcal Z_\alpha.
\end{equation}
\end{subequations}
Eq.~\eqref{spG} can be solved numerically for a given value of $\alpha$, allowing then to evaluate $\alpha\Phi(\alpha)$ in Eq.~\eqref{aPhi}, whose Legendre transform is the desired large deviation function $L(C)$. Using the properties of $G_\alpha$ derived above, it can be seen that $\lim_{\alpha\to -2^+}\alpha\Phi(\alpha)=-\infty$: the presence of such a singularity gives us information on the large $C$ behavior of the Cram\'er function, i.e., it implies that $\lim_{C\to+\infty}C^{-1}L(C)=2$. As anticipated, Eqs.~\eqref{PhiG} have been derived assuming $-2< \alpha<0$. To get an expression that can be prolonged to positive values of $\alpha$ we will use the cavity method.

\section{Large deviations via cavity} The equation for $\Phi(\alpha)$ given by the replica method for $\alpha\in(-2,0]$ can be also obtained using the cavity method and actually extended to positive values of $\alpha$. The starting point is the cavity condition for the occupancy of an edge in the random-link matching problem. In particular, in the cavity approach, each edge $(i,j)$ is associated to its weight $w_{ij}$ and to two cavity fields, $\phi_{i}$ and $\phi_{j}$ on its vertices, containing information on the rest of the graph, in such a way that the occupancy $m_{ij}$ of the the edge is distributed as 
\begin{equation}
 P(m_{ij})=\frac{\exp\left[-\beta m_{ij}(Nw_{ij}-\phi_{i}-\phi_{j})\right]}{1+\exp\left[-\beta (Nw_{ij}-\phi_{i}-\phi_{j})\right]}.\label{pm}
\end{equation}
In the $\beta\to+\infty$ limit, an edge is occupied if, and only if, $Nw_{ij}<\phi_i+\phi_j$. At zero temperature and in the large $N$ limit, the cavity fields satisfy the following equation \cite{Mezard1986a,Krauth1989a,Parisi2001}
\begin{equation}
 \phi_{0}=\min_{k\in\partial 0}\left(Nw_{k0}-\phi_{k}\right).\label{cavita}
\end{equation}
Here $\partial 0$ is the set of neighbors of the node $0$. The mate node $i^*$ of $0$ is such that
\begin{equation}
 i^*=\arg\min_{k\in\partial 0}\left(Nw_{k0}-\phi_{k}\right)
\end{equation}
In Refs.~\cite{Mezard1986a,Krauth1989a,Parisi2001} the previous equations have been studied and solved. The AOC predicted by the cavity method coincides with the one obtained using the replica approach. 

The recurrence relation for the cavity fields can be also used to extract information on the fluctuations and evaluate $\alpha\Phi(\alpha)$. For the sake of simplicity, let us start from a different version of the problem, i.e., the random-link matching problem on a sparse graph, and in particular on a Bethe lattice, and let us follow the approach of \textcite{Rivoire2005} for the study of large deviation on sparse topologies. In this case, we are interested in solving our problem on a graph having $2N$ vertices, each one of them having coordination $z$: we will later take the limit $z\to 2N-1$. Taking this limit might sound dangerous, because we apply a result obtained for a sparse topology to a dense one. However, the random-link matching problem is an ``effectively sparse'' problem: given a fully-connected topology, the probability that a given node is connected, in the optimal matching, to its $n$th nearest neighbor is exponentially small in $n$ \cite{Parisi2001}. We will denote by $L_z(C)$ the large deviation function for random-link matching problem on the Bethe lattice, so that $L(C)=\lim_{z\to+\infty}L_z(C)$. 

To obtain an expression for it, we proceed as usual in the cavity approach, i.e., starting from an intermediate graph having $z$ randomly chosen (cavity) nodes with coordination $z-1$, and all the other nodes with coordination $z$.
\begin{center}
\begin{tikzpicture}
  \draw (0,0) node[disc=c-0-1,draw=none];
  \xdef\radius{0cm}
  \xdef\level{0}
  \xdef\nbnodes{1}
  \xdef\degree{5} 
  \pgfmathsetmacro\nlevel{int(\level+1)}
    \pgfmathsetmacro\nnbnodes{int(\nbnodes*(\degree-1))}
    \pgfmathsetmacro\nradius{\radius+0.4cm}
    \foreach \div in {1,...,\nnbnodes} {
      \pgfmathtruncatemacro\src{((\div+\degree-2)/(\degree-1))}
      \path (c-0-1) ++({\div*(360/\nnbnodes)-180/\nnbnodes}:\nradius pt) node[disc=c-\nlevel-\div,fill=red];
    }
    \xdef\radius{\nradius}
    \xdef\level{\nlevel}
    \xdef\nbnodes{\nnbnodes}
    \xdef\degree{4}
  \foreach \ndegree/\form in {4/disc,4/disc,4/disc}{
    \pgfmathsetmacro\nlevel{int(\level+1)}
    \pgfmathsetmacro\nnbnodes{int(\nbnodes*(\degree-1))}
    \pgfmathsetmacro\nradius{\radius+0.4cm}
    \foreach \div in {1,...,\nnbnodes} {
      \pgfmathtruncatemacro\src{((\div+\degree-2)/(\degree-1))}
      \path (c-0-1) ++({\div*(360/\nnbnodes)-180/\nnbnodes}:\nradius pt) node[\form=c-\nlevel-\div];
      \draw (c-\level-\src) -- (c-\nlevel-\div);
    }
    \xdef\radius{\nradius}
    \xdef\level{\nlevel}
    \xdef\nbnodes{\nnbnodes}
    \xdef\degree{\ndegree}
  }
\end{tikzpicture}    
\end{center}
Let us denote by $\hat L_z$ the large deviation function corresponding to the random-link matching problem on such a topology. 

We can recover the ``correct'' Bethe lattice topology in two ways. We can, for example, connect a new node to the $z$ cavity nodes.
        \begin{center}
\begin{tikzpicture}
  \draw (0,0) node[disc=c-0-1,fill=red];
  \xdef\radius{0cm}
  \xdef\level{0}
  \xdef\nbnodes{1}
  \xdef\degree{5} 
  \foreach \ndegree/\form in {4/disc,4/disc,4/disc,4/disc}{
    \pgfmathsetmacro\nlevel{int(\level+1)}
    \pgfmathsetmacro\nnbnodes{int(\nbnodes*(\degree-1))}
    \pgfmathsetmacro\nradius{\radius+0.4cm}
    \foreach \div in {1,...,\nnbnodes} {
      \pgfmathtruncatemacro\src{((\div+\degree-2)/(\degree-1))}
      \path (c-0-1) ++({\div*(360/\nnbnodes)-180/\nnbnodes}:\nradius pt) node[\form=c-\nlevel-\div];
      \draw (c-\level-\src) -- (c-\nlevel-\div);
    }
    \xdef\radius{\nradius}
    \xdef\level{\nlevel}
    \xdef\nbnodes{\nnbnodes}
    \xdef\degree{\ndegree}
  }
\end{tikzpicture}    
\end{center}
The optimal cost will be shifted by a certain amount $N^{-1}\varepsilon$, in such a way that the probability density function of the optimal cost satisfies the equation
\begin{multline}\label{ldcavita0}
    \e^{-(N+\sfrac{1}{2})L_{z}(C)}
    =\int \e^{-N\hat L_z\left(\frac{N+\sfrac{1}{2}}{N}C-\varepsilon\right)} p_v(\varepsilon)\dd\varepsilon\\
    \asymp \e^{-N \hat L_z(C)+\frac{\alpha}{2}C}\int\e^{-\alpha N\varepsilon } p_v(\varepsilon)\dd\varepsilon,
\end{multline}
where $\alpha\coloneqq -\partial_C \hat L_z(C)$ and $p_v(\varepsilon)$ is the distribution of the energy-shift $\varepsilon$ due to a vertex addition. 

Another possibility is to add $\sfrac{z}{2}$ edges.
        \begin{center}
\begin{tikzpicture}
  \draw (0,0) node[disc=c-0-1,draw=none];
  \xdef\radius{0cm}
  \xdef\level{0}
  \xdef\nbnodes{1}
  \xdef\degree{5} 
  \pgfmathsetmacro\nlevel{int(\level+1)}
    \pgfmathsetmacro\nnbnodes{int(\nbnodes*(\degree-1))}
    \pgfmathsetmacro\nradius{\radius+0.4cm}
    \foreach \div in {1,...,\nnbnodes} {
      \pgfmathtruncatemacro\src{((\div+\degree-2)/(\degree-1))}
      \path (c-0-1) ++({\div*(360/\nnbnodes)-180/\nnbnodes}:\nradius pt) node[disc=c-\nlevel-\div,fill=gray];
    }
    \draw [red] (c-1-2) -- (c-1-1);\draw [red] (c-1-3) -- (c-1-4);
    \xdef\radius{\nradius}
    \xdef\level{\nlevel}
    \xdef\nbnodes{\nnbnodes}
    \xdef\degree{4}
  \foreach \ndegree/\form in {4/disc,4/disc,4/disc}{
    \pgfmathsetmacro\nlevel{int(\level+1)}
    \pgfmathsetmacro\nnbnodes{int(\nbnodes*(\degree-1))}
    \pgfmathsetmacro\nradius{\radius+0.4cm}
    \foreach \div in {1,...,\nnbnodes} {
      \pgfmathtruncatemacro\src{((\div+\degree-2)/(\degree-1))}
      \path (c-0-1) ++({\div*(360/\nnbnodes)-180/\nnbnodes}:\nradius pt) node[\form=c-\nlevel-\div];
      \draw (c-\level-\src) -- (c-\nlevel-\div);
    }
    \xdef\radius{\nradius}
    \xdef\level{\nlevel}
    \xdef\nbnodes{\nnbnodes}
    \xdef\degree{\ndegree}
  }
\end{tikzpicture}    
\end{center}
We obtain in this case the following relation
\begin{multline}\label{ldcavita1}
    \e^{-NL_{z}(C)}=\left(\prod_{k=1}^{\sfrac{z}{2}} \int p_e(\epsilon_k)\dd\epsilon_k \right)\e^{-N\hat L_z\left(C-\sum_k\epsilon_k\right)}\\\asymp\e^{-N\hat L_z(C)}\left(\int p_e(\epsilon)\e^{-\alpha N\epsilon}\dd\epsilon\right)^\frac{z}{2},
\end{multline}
where $p_e(\epsilon)$ is the distribution of the energy-shift $\varepsilon$ due to an edge addition. Taking the ratio of the two expressions above, we obtain an equation for $L_z(C)$ at the leading order in $N$, namely
\begin{multline}
    \alpha\Phi(\alpha)\equiv L_z(C)+\alpha C\\=\lim_N\!\left[z\ln\left(\!\int p_e(\epsilon)\e^{-\alpha N\epsilon}\dd\epsilon\!\right)\!-\!2\ln\left(\!\int\e^{-\alpha N\varepsilon } p_v(\varepsilon)\dd\varepsilon\!\right)\right].
\end{multline}
Taking $z=2N-1\approx 2N$ we obtain the expression for our case. To evaluate the previous quantity, let us introduce the joint distribution $p_v(\phi,\varepsilon)$ of the cavity field entering in the added node and of the energy shift, such that $p_v(\varepsilon)=\int p_v(\phi,\varepsilon)\dd\phi$. We define the reweighted distribution of the cavity field 
\begin{equation}
    p_\alpha(\phi)\coloneqq\frac{1}{\mathcal Z_\alpha}\int \e^{-\alpha N\varepsilon } p_v(\phi,\varepsilon)\dd\varepsilon,
\end{equation}
with $\mathcal Z_\alpha$ proper normalization constant. In our case \[p_v(\phi,\varepsilon)\equiv p_v(\varepsilon)\delta\left(\frac{\phi}{N}-\varepsilon\right),\] because the cost shift due to the addition of a node coincides with the incoming cavity field, see Eq.~\eqref{cavita}. If we denote by \[\pi_\alpha(u)\coloneqq \int_0^{+\infty}p_\alpha(\hat w-u)\dd\hat w,\] the distribution of the cavity field can be rewritten as
\begin{multline}
p_\alpha(\phi)=\frac{1}{\mathcal Z_\alpha} \int p_v(\phi,\varepsilon)\e^{-\alpha N\varepsilon}\dd\varepsilon=\frac{p_v(\sfrac{\phi}{N})\e^{-\alpha\phi}}{\mathcal Z_\alpha}\\
= \frac{2\e^{-\alpha\phi}\pi_\alpha(\phi)}{\mathcal Z_\alpha}\left(1-\!\frac{1}{N}\int^{\phi}_{-\infty}\pi_\alpha(\chi)\dd\chi\right)^{\!\!2N-1},
\end{multline}
where we have used Eq.~\eqref{cavita} to express $p_v$ in terms of $\pi_\alpha$. In the $N\to +\infty$ limit we obtain an equation for $p_\alpha$, 
\begin{equation}\label{p-pi}
p_\alpha(\phi)
= \frac{2\e^{-\alpha\phi}\pi_\alpha(\phi)}{\mathcal Z_\alpha}\exp\left(-2\int^{\phi}_{-\infty}\pi_\alpha(\chi)\dd\chi\right).
\end{equation}
Moreover, because of Eq.~\eqref{pm}, the energy cost due to the addition of an edge
\begin{multline}
    p_e(\epsilon)=\iint \dd\phi_1 \dd\phi_2\,p_\alpha(\phi_1) p_\alpha(\phi_2)\\\times\int_{0}^{+\infty}\dd w\,\rho(w)\delta\left(\epsilon-\min\left\{0,w-\frac{\phi_1-\phi_2}{N}\right\}\right),
\end{multline}
and therefore, by means of an integration by parts, we get for large $N$
\begin{multline}
 \alpha\Phi(\alpha)\coloneqq \alpha C+L(C)
 =-2\ln\mathcal Z_\alpha\\+\frac{\alpha}{\mathcal Z_\alpha}\iint \dd\phi_1 \dd\phi_2\,p_\alpha(\phi_1) p_\alpha(\phi_2)\\
 \times \int_{0}^{\infty} \dd w\,w\e^{-\alpha(w-\phi_1-\phi_2)}\theta(\phi_1+\phi_2-w).\label{aPhicav}
\end{multline}
Note that, up to now, no assumptions have been made on the range of values of $\alpha$, except the implicit ones about the fact that the quantities above are well defined and convergent: the cavity expression can be used therefore for both positive and negative values of $\alpha$, provided that the involved quantities are finite. 

It can be seen that the expression in Eq.~\eqref{aPhicav} is equivalent to the one given in Eq.~\eqref{aPhi} in its range of validity.
Indeed, introducing the function
\begin{equation}\label{defg}
 G_\alpha(\phi)\coloneqq 2\int_0^{+\infty}\hat w p_\alpha(\hat w-\phi)\dd\hat w,
\end{equation}
Eq.~\eqref{p-pi} simplifies as
\begin{equation}
 p_\alpha(\phi)=\frac{1}{\mathcal Z_\alpha}\frac{\dd G_\alpha(\phi)}{\dd\phi}\e^{-\alpha\phi-G_\alpha(\phi)},
 \label{pG}
\end{equation}
and therefore we can write, using Eq.~\eqref{defg}, a self-consistent equation for the function $G_{\alpha}(\phi)$ that is found to be identical to Eq.~\eqref{spG}, proving that the function $G_\alpha$ introduced here is the same appearing in the replica approach. Repeating the arguments presented in the replica derivation, we obtain that the integrals are finite for $\alpha> -2$. Substituting Eq.~\eqref{pG} in Eq.~\eqref{aPhicav}, simple manipulations give us the same expression presented in Eq.~\eqref{aPhi}, the main difference being that, in the obtained formula,
\begin{equation}
    \mathcal Z_\alpha=\int \e^{-\alpha\phi}p_v(\phi)\dd\phi=\int \e^{-\alpha\phi}\frac{\dd G_\alpha(\phi)}{\dd\phi}\e^{-G_\alpha(\phi)}\dd\phi.
\end{equation}
If we further restrict ourselves to negative values of $\alpha$, an integration by parts allows us to write $\mathcal Z_\alpha$ in the same form given in Eq.~\eqref{spG}, proving the equivalence of the two approaches and implicitly providing us a prolongation of the replica result.

\section{Numerical results}
\begin{figure}
 \includegraphics[width=\columnwidth]{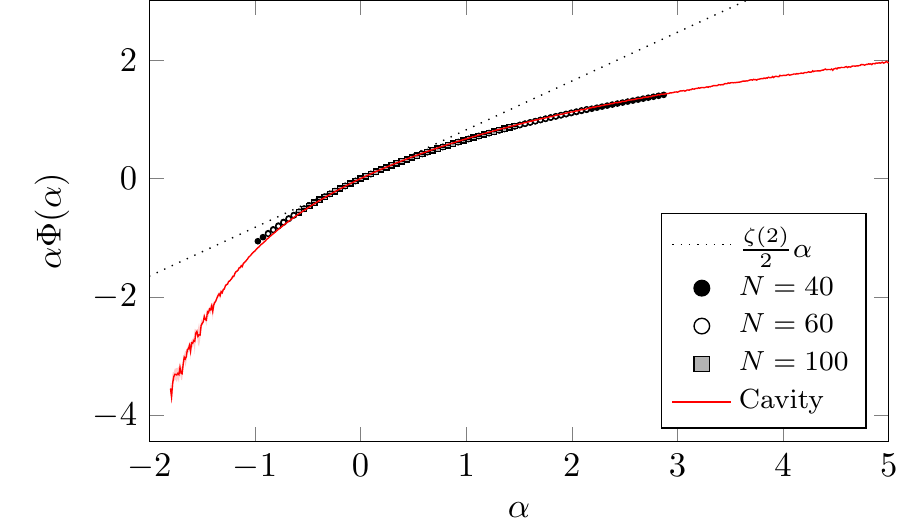}
 \caption{Cavity and numerical results for the $\alpha\Phi(\alpha)$ in the random-link matching problem. The cavity results have been obtained by means of a population dynamics algorithm, using a population of $10^5$ fields. The numerical data have been obtained instead from $10^{10}$ instances for each value of $\alpha$ and $N$: for large $N$ the finiteness of the number of instances makes the curves going to $\alpha\sfrac{\zeta(2)}{2}$ because of the concentration of the measure. \label{fig:Phi}}
\end{figure}

\begin{figure}
\includegraphics[width=\columnwidth]{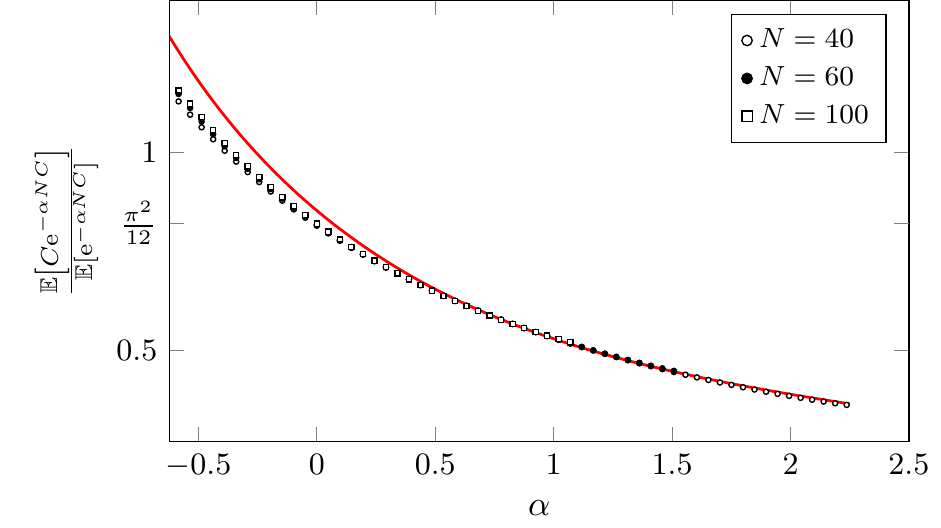}
 \caption{Plot of the derivative of $\alpha\Phi(\alpha)$ obtained from numerical simulation compared with the theoretical prediction obtained using cavity.\label{fig:Phi1}}
\end{figure}
\begin{figure}
    \includegraphics[width=\columnwidth]{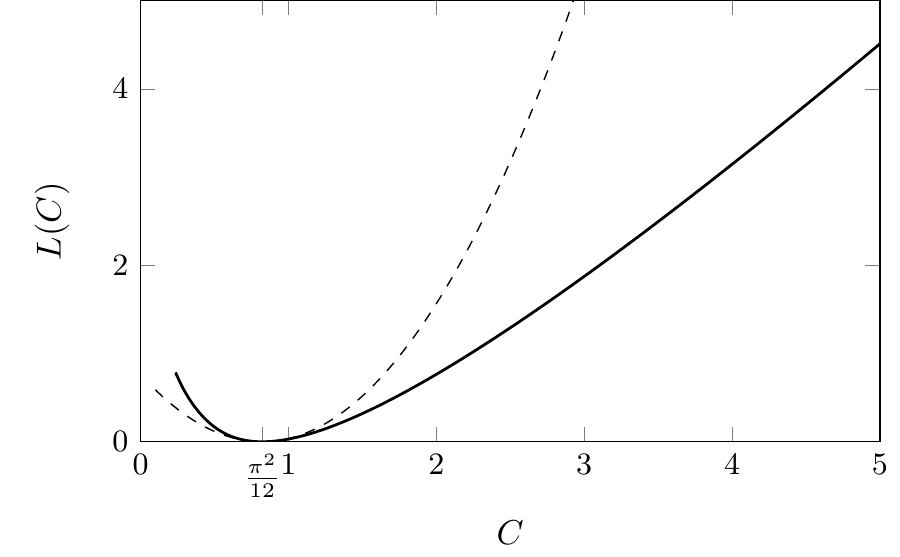}
    \caption{Function $L(C)$ obtained performing the Legendre transform of our cavity results, compared with the quadratic approximation (dashed) provided by the results in Section~\ref{sec:small}.}
    \label{fig:Phi2}
 \end{figure}

We have integrated Eq.~\eqref{aPhicav} by means of a population dynamics algorithm, and we have compared our results with the value of $\alpha\Phi(\alpha)$ obtained by numerical simulations, performed solving a large number of instances of the problem. In our numerical simulations, we assumed $\rho(w)=\e^{-w}$. The quantity $\alpha\Phi(\alpha)$ can be evaluated directly, using the fact that $\alpha\Phi(\alpha)=\lim_{N\to+\infty} N^{-1}\ln\mathbb E\left[\e^{-\alpha N C_N}\right]$. The agreement is very good in the neighborhood of the origin, as it can be seen in Fig.~\ref{fig:Phi}, where the cavity and numerical results are compared. For reference, we have plotted the tangent to the curve in the origin, whose slope coincides with the AOC, $\sfrac{\zeta(2)}{2}$. We have discarded the data points where finite-sample effects appear for larger values of $|\alpha|$; obviously, better estimates can be obtained running an exponentially large number of instances in the size of the system. Finite-size effects still appear in the evaluation of the derivative of $\alpha\Phi(\alpha)$ (that has been also directly evaluated as $\partial_\alpha[\alpha\Phi(\alpha)]=\lim_{N\to+\infty}\frac{\mathbb E[C_N\e^{-\alpha N C_N}]}{\mathbb E[\e^{-\alpha N C_N}]}$) for $\alpha<0$, larger sizes being closer to the theoretical predictions, see Fig.~\ref{fig:Phi1}. Finally, the theoretical prediction near $\alpha=-2$ becomes noisy and less reliable, because of the approaching of the divergence. In this regime, the convergence of the cavity fields distribution fails and an accurate estimation of $\alpha\Phi(\alpha)$ is more challenging. In Fig.~\ref{fig:Phi2} we have plotted the Legendre transform of our cavity results, that is the desired large deviation function. The curve is compared with the quadratic approximation in the origin, corresponding to the Gaussian small-fluctuation behavior.
\section{Conclusions}
Using the replica approach we have evaluated the variance of the average optimal cost for the random-link matching problem on the complete graph, assuming a distribution of the weights such that $\lim_{w\to 0}\rho(w)=1$. Our result is in agreement with a previously obtained expression by W\"{a}stlund, proving that the small fluctuations of the optimal cost around its asymptotic AOC are Gaussian. We have then derived an expression for the Legendre transform $\alpha\Phi(\alpha)$ of the Cram\'er function $L(C)$ using both the replica theory (for positive cost fluctuation) and the cavity method. The cavity formula, in particular, has been obtained for the random-link matching problem on a generic sparse graph having fixed coordination, and provides a recipe for the numerical evaluation of $\alpha\Phi(\alpha)$. In the fully-connected case, our results also show that $\alpha\Phi(\alpha)$ diverges for $\alpha\to-2$, implying that $\lim_{C\to+\infty}C^{-1}L(C)=2$.

\subsection*{Acknowledgments} We thank Johan W\"{a}stlund for communicating us his results. GS also thanks Sergio Caracciolo and Yan Fyodorov for stimulating the study of large deviations in the random-link matching problem. We thank Federico Ricci-Tersenghi and Tommaso Rizzo for useful discussions. GP and GS acknowledge financial support from the Simons Foundation (grant No. 454949, G.~Parisi).

\appendix*
\section{Derivation of Eq.~\eqref{limite}}
To prove the limit in Eq.~\eqref{limite}, we can start using the fact that $\binom{n}{k}=(-1)^k\binom{-n+k-1}{k}$. The series can be written as
\begin{equation}
 \sum_{p=1}^\infty  \frac{1}{\binom{\sfrac{\alpha}{\beta}}{p}}\frac{\e^{\beta u p}}{(p!)^2}=\frac{\Gamma\left(-\sfrac{\alpha}{\beta}\right)}{\beta}\sum_{p=1}^\infty  \frac{1}{\Gamma\left(p-\sfrac{\alpha}{\beta}\right)}\frac{(-\e^{\beta u })^p}{p!p}.
\end{equation}
We have to compute the $\beta\to+\infty$ limit at $n\beta=\alpha$ fixed. The coefficient in front of the series has limit $\lim_{\beta\to+\infty}\beta^{-1}\Gamma\left(-\sfrac{\alpha}{\beta}\right)=-\sfrac{1}{\alpha}$. On the other hand,
\begin{multline}
 \lim_{\beta\to+\infty}\sum_{p=1}^\infty  \frac{1}{\Gamma\left(p-\sfrac{\alpha}{\beta}\right)}\frac{(-\e^{\beta u })^p}{p!p}+\theta(u)\\
 =\lim_{\beta\to+\infty}\sum_{p=1}^\infty  \left(\frac{\Gamma(p)}{\Gamma\left(p-\sfrac{\alpha}{\beta}\right)}-1\right)\frac{(-\e^{\beta u })^p}{(p!)^2},
\end{multline} 
where we have used the fact that
\begin{equation}
 \lim_{\beta\to+\infty}\sum_{p=1}^\infty\frac{(-\e^{\beta u })^p}{(p!)^2}=-\theta(u).
\end{equation}
We can write the following expansion
\begin{multline} \label{bellexp}
\frac{\Gamma(p)}{\Gamma\left(p-\sfrac{\alpha}{\beta}\right)}-1\\
=\sum_{k=1}^\infty\frac{\left(-\sfrac{\alpha}{\beta}\right)^k}{k!}\mathrm B_{k}\left(\hat\psi_0(p),\hat\psi_1(p),\dots,\hat\psi_{k-1}(p)\right),
\end{multline} 
where we have introduced the (opposite) polygamma function $\hat\psi_n(z)\coloneqq -\psi_n(z)$
\begin{equation}\label{polygamma}
\psi_n(z)\coloneqq \frac{\dd^{n+1}}{\dd z^{n+1}}\ln\Gamma(z)=(-1)^{n+1}\int_{0}^\infty t^{n}\frac{\e^{-tz}}{1-\e^{-t}}\dd t
\end{equation}
and the (complete) exponential Bell polynomials 
\begin{equation}
 \mathrm B_{k}(x_1,\dots,x_{k})\coloneqq \left.\left(\frac{\partial}{\partial t}\right)^k\exp\left(\sum_{j=1}^\infty\frac{x_jt^j}{j!}\right)\right|_{t=0}.
\end{equation}
The expansion above is obtained applying the Fa\`a di Bruno formula for the derivative of a composed function $\partial^n_x (f\circ g)(x)$ with $f(x)\equiv \e^{x}$ and $g(x)\equiv -\ln\Gamma(x)$. If we now assume $\mathrm B_0\equiv 1$, the Bell polynomials satisfy the recurrence relations \cite{Charalambides2002}
\begin{equation}
 \mathrm B_{k+1}(x_1,\dots,x_{k+1})=\sum_{i=0}^k\binom{k}{i}\mathrm B_{k-i}(x_1,\dots,x_{k-i})x_{i+1}.
\end{equation}
We will show now by induction that, for $k\geq 2$,
\begin{equation}
\lim_{\beta\to+\infty}\sum_{p=1}^\infty\frac{\mathrm B_{k}(\hat\psi_0(p),\dots,\hat\psi_{k-1}(p))}{\beta^k}\frac{(-\e^{\beta u })^p}{(p!)^2}=0.\end{equation}
For $k=1$ we have
\begin{multline} 
 \lim_{\beta\to+\infty}\sum_{p=1}^\infty\frac{\hat\psi_0(p)}{\beta}\frac{(-\e^{\beta u })^p}{(p!)^2}\\
 =\lim_{\beta\to+\infty}\int_0^{+\infty}\frac{J_0(2\e^{\beta\frac{u-t}{2}})-1}{1-\e^{-\beta t}}\dd t=-u\theta(u).
\end{multline} 
The $k=2$ case follows straightforwardly. Indeed,
\begin{multline}
 \lim_{\beta\to+\infty}\sum_{p=1}^\infty\frac{\mathrm B_{2}(\hat\psi_0(p),\hat\psi_{1}(p))}{\beta^2}\frac{(-\e^{\beta u })^p}{(p!)^2}\\
 =\lim_{\beta\to+\infty}\sum_{p=1}^\infty\frac{\hat\psi_0^2(p)+\hat\psi_{1}(p)}{\beta^2}\frac{(-\e^{\beta u })^p}{(p!)^2}\\
 =-\int_0^{+\infty}\dd t\int_0^{+\infty}\dd\tau\,\theta(u-t-\tau)+\int_0^{+\infty}\dd t \, t\theta(u-t)=0.
\end{multline}
Let us suppose now that the statement is true for generic $k>2$ and let us consider it for $k+1$. Then we have
\begin{widetext}
\begin{equation} 
 \sum_{p=1}^\infty\frac{\mathrm B_{k+1}(\hat\psi_0(p),\dots,\hat\psi_{k}(p))}{\beta^k}\frac{(-\e^{\beta u })^p}{(p!)^2}=\sum_{i=0}^k\binom{k}{i}\sum_{p=1}^\infty\frac{\mathrm B_{k-i}(\hat\psi_0(p),\dots,\hat\psi_{k-i-1}(p))\hat\psi_{i}(p)}{\beta^{k+1}}\frac{(-\e^{\beta u })^p}{(p!)^2}.
\end{equation} 
The inner sum can be written as
\begin{equation} 
 \int_{0}^{+\infty} \dd t \, \frac{(-1)^it^i}{1-\e^{-\beta t}}\sum_{p=1}^\infty\frac{\mathrm B_{k-i}(\hat\psi_0(p),\dots,\hat\psi_{k-i-1}(p))}{\beta^{k-i}}\frac{(-\e^{\beta (u-t) })^p}{(p!)^2}
\end{equation} 
implying that
\begin{multline}
 \sum_{p=1}^\infty\frac{\mathrm B_{k+1}(\hat\psi_0(p),\dots,\hat\psi_{k}(p))}{\beta^k}\frac{(-\e^{\beta u })^p}{(p!)^2}=\sum_{i=0}^{k-2}\binom{k}{i}\int_{0}^{+\infty}\dd t \, \frac{(-1)^it^i}{1-\e^{-\beta t}}\sum_{p=1}^\infty\frac{\mathrm B_{k-i}(\hat\psi_0(p),\dots,\hat\psi_{k-i-1}(p))}{\beta^{k-i}}\frac{(-\e^{\beta (u-t) })^p}{(p!)^2}\\
 +k(-1)^{k-1} \int_{0}^{+\infty}\dd t \, \frac{t^{k-1}}{1-\e^{-\beta t}}\sum_{p=1}^\infty \frac{\hat\psi_0(p)}{\beta} \frac{(-\e^{\beta (u-t) })^p}{(p!)^2}+ (-1)^k \int_{0}^{+\infty}\dd t \, \frac{t^k}{1-\e^{-\beta t}}\sum_{p=1}^\infty\frac{(-\e^{\beta (u-t) })^p}{(p!)^2}.
\end{multline}
\end{widetext}
The last two terms in the previous expression in the $\beta \to \infty$ tend to zero
\begin{multline}
-k(-1)^{k-1}\iint_{0}^{+\infty} t^{k-1}\theta(u-t-\tau)\dd\tau\dd t \\+ (-1)^{k-1} \int_{0}^{+\infty} \dd t \, t^k \theta(u-t) = 0
\end{multline}
By the induction hypothesis, the remaining $k-2$ contributions are infinitesimal as well for $\beta\to+\infty$, and the thesis is proved. We can restrict therefore the expansion in Eq.~\eqref{bellexp} to the $k=1$ term in the $\beta\to+\infty$ hypothesis. We have
\begin{equation}
    \lim_{\beta\to+\infty}\sum_{p=1}^\infty  \left(\frac{\Gamma(p)}{\Gamma\left(p-\sfrac{\alpha}{\beta}\right)}-1\right)\frac{(-\e^{\beta u })^p}{(p!)^2} =\alpha u\theta(u) .
\end{equation}
The relations above finally give us the asymptotic in Eq.~\eqref{limite}.
\bibliography{biblio.bib}

\begin{thebibliography}{33}%
\makeatletter
\providecommand \@ifxundefined [1]{%
 \@ifx{#1\undefined}
}%
\providecommand \@ifnum [1]{%
 \ifnum #1\expandafter \@firstoftwo
 \else \expandafter \@secondoftwo
 \fi
}%
\providecommand \@ifx [1]{%
 \ifx #1\expandafter \@firstoftwo
 \else \expandafter \@secondoftwo
 \fi
}%
\providecommand \natexlab [1]{#1}%
\providecommand \enquote  [1]{``#1''}%
\providecommand \bibnamefont  [1]{#1}%
\providecommand \bibfnamefont [1]{#1}%
\providecommand \citenamefont [1]{#1}%
\providecommand \href@noop [0]{\@secondoftwo}%
\providecommand \href [0]{\begingroup \@sanitize@url \@href}%
\providecommand \@href[1]{\@@startlink{#1}\@@href}%
\providecommand \@@href[1]{\endgroup#1\@@endlink}%
\providecommand \@sanitize@url [0]{\catcode `\\12\catcode `\$12\catcode
  `\&12\catcode `\#12\catcode `\^12\catcode `\_12\catcode `\%12\relax}%
\providecommand \@@startlink[1]{}%
\providecommand \@@endlink[0]{}%
\providecommand \url  [0]{\begingroup\@sanitize@url \@url }%
\providecommand \@url [1]{\endgroup\@href {#1}{\urlprefix }}%
\providecommand \urlprefix  [0]{URL }%
\providecommand \Eprint [0]{\href }%
\providecommand \doibase [0]{http://dx.doi.org/}%
\providecommand \selectlanguage [0]{\@gobble}%
\providecommand \bibinfo  [0]{\@secondoftwo}%
\providecommand \bibfield  [0]{\@secondoftwo}%
\providecommand \translation [1]{[#1]}%
\providecommand \BibitemOpen [0]{}%
\providecommand \bibitemStop [0]{}%
\providecommand \bibitemNoStop [0]{.\EOS\space}%
\providecommand \EOS [0]{\spacefactor3000\relax}%
\providecommand \BibitemShut  [1]{\csname bibitem#1\endcsname}%
\let\auto@bib@innerbib\@empty
\bibitem [{\citenamefont {M{\'e}zard}\ \emph {et~al.}(1987)\citenamefont
  {M{\'e}zard}, \citenamefont {Parisi},\ and\ \citenamefont
  {Virasoro}}]{mezard1987spin}%
  \BibitemOpen
  \bibfield  {author} {\bibinfo {author} {\bibfnamefont {M.}~\bibnamefont
  {M{\'e}zard}}, \bibinfo {author} {\bibfnamefont {G.}~\bibnamefont {Parisi}},
  \ and\ \bibinfo {author} {\bibfnamefont {M.}~\bibnamefont {Virasoro}},\
  }\href {http://books.google.it/books?id=ZIF9QgAACAAJ} {\emph {\bibinfo
  {title} {Spin Glass Theory and Beyond}}},\ Lecture Notes in Physics Series\
  (\bibinfo  {publisher} {World Scientific Publishing Company, Incorporated},\
  \bibinfo {year} {1987})\BibitemShut {NoStop}%
\bibitem [{\citenamefont {Orland}(1985)}]{Orland1985}%
  \BibitemOpen
  \bibfield  {author} {\bibinfo {author} {\bibfnamefont {H.}~\bibnamefont
  {Orland}},\ }\href {\doibase 10.1051/jphyslet:019850046017076300} {\bibfield
  {journal} {\bibinfo  {journal} {J. Phys. Lett. (Paris)}\ }\textbf {\bibinfo
  {volume} {46}},\ \bibinfo {pages} {763} (\bibinfo {year} {1985})}\BibitemShut
  {NoStop}%
\bibitem [{\citenamefont {M\'{e}zard}\ and\ \citenamefont
  {Parisi}(1985)}]{Mezard1985}%
  \BibitemOpen
  \bibfield  {author} {\bibinfo {author} {\bibfnamefont {M.}~\bibnamefont
  {M\'{e}zard}}\ and\ \bibinfo {author} {\bibfnamefont {G.}~\bibnamefont
  {Parisi}},\ }\href {\doibase 10.1051/jphyslet:019850046017077100} {\bibfield
  {journal} {\bibinfo  {journal} {J. Phys. Lett. (Paris)}\ }\textbf {\bibinfo
  {volume} {46}},\ \bibinfo {pages} {771} (\bibinfo {year} {1985})}\BibitemShut
  {NoStop}%
\bibitem [{\citenamefont {M\'{e}zard}\ and\ \citenamefont
  {Parisi}(1988)}]{Mezard1988}%
  \BibitemOpen
  \bibfield  {author} {\bibinfo {author} {\bibfnamefont {M.}~\bibnamefont
  {M\'{e}zard}}\ and\ \bibinfo {author} {\bibfnamefont {G.}~\bibnamefont
  {Parisi}},\ }\href {\doibase 10.1051/jphys:0198800490120201900} {\bibfield
  {journal} {\bibinfo  {journal} {J. Phys. (Paris)}\ }\textbf {\bibinfo
  {volume} {49}},\ \bibinfo {pages} {2019} (\bibinfo {year}
  {1988})}\BibitemShut {NoStop}%
\bibitem [{\citenamefont {Lucibello}\ \emph {et~al.}(2017)\citenamefont
  {Lucibello}, \citenamefont {Parisi},\ and\ \citenamefont
  {Sicuro}}]{Lucibello2017}%
  \BibitemOpen
  \bibfield  {author} {\bibinfo {author} {\bibfnamefont {C.}~\bibnamefont
  {Lucibello}}, \bibinfo {author} {\bibfnamefont {G.}~\bibnamefont {Parisi}}, \
  and\ \bibinfo {author} {\bibfnamefont {G.}~\bibnamefont {Sicuro}},\ }\href
  {\doibase 10.1103/PhysRevE.95.012302} {\bibfield  {journal} {\bibinfo
  {journal} {Phys. Rev. E}\ }\textbf {\bibinfo {volume} {95}},\ \bibinfo
  {pages} {012302} (\bibinfo {year} {2017})}\BibitemShut {NoStop}%
\bibitem [{\citenamefont {Lucibello}\ \emph {et~al.}(2018)\citenamefont
  {Lucibello}, \citenamefont {Malatesta}, \citenamefont {Parisi},\ and\
  \citenamefont {Sicuro}}]{Lucibello2018}%
  \BibitemOpen
  \bibfield  {author} {\bibinfo {author} {\bibfnamefont {C.}~\bibnamefont
  {Lucibello}}, \bibinfo {author} {\bibfnamefont {E.~M.}\ \bibnamefont
  {Malatesta}}, \bibinfo {author} {\bibfnamefont {G.}~\bibnamefont {Parisi}}, \
  and\ \bibinfo {author} {\bibfnamefont {G.}~\bibnamefont {Sicuro}},\ }\href
  {http://stacks.iop.org/1742-5468/2018/i=5/a=053301} {\bibfield  {journal}
  {\bibinfo  {journal} {J. Stat. Mech. Theory Exp.}\ }\textbf {\bibinfo
  {volume} {2018}},\ \bibinfo {pages} {053301} (\bibinfo {year}
  {2018})}\BibitemShut {NoStop}%
\bibitem [{\citenamefont {M\'{e}zard}\ and\ \citenamefont
  {Parisi}(1986{\natexlab{a}})}]{Mezard1986}%
  \BibitemOpen
  \bibfield  {author} {\bibinfo {author} {\bibfnamefont {M.}~\bibnamefont
  {M\'{e}zard}}\ and\ \bibinfo {author} {\bibfnamefont {G.}~\bibnamefont
  {Parisi}},\ }\href
  {http://jphys.journaldephysique.org/index.php?option=com\_article\&access=doi\&doi=10.1051/jphys:019860047080128500}
  {\bibfield  {journal} {\bibinfo  {journal} {J. Phys. (Paris)}\ }\textbf
  {\bibinfo {volume} {47}},\ \bibinfo {pages} {1285} (\bibinfo {year}
  {1986}{\natexlab{a}})}\BibitemShut {NoStop}%
\bibitem [{\citenamefont {Krauth}\ and\ \citenamefont
  {M\'{e}zard}(1989)}]{Krauth1989a}%
  \BibitemOpen
  \bibfield  {author} {\bibinfo {author} {\bibfnamefont {W.}~\bibnamefont
  {Krauth}}\ and\ \bibinfo {author} {\bibfnamefont {M.}~\bibnamefont
  {M\'{e}zard}},\ }\href@noop {} {\bibfield  {journal} {\bibinfo  {journal}
  {Europhysics Letters}\ }\textbf {\bibinfo {volume} {8}},\ \bibinfo {pages}
  {213} (\bibinfo {year} {1989})}\BibitemShut {NoStop}%
\bibitem [{\citenamefont {M{\'e}zard}\ \emph {et~al.}(2002)\citenamefont
  {M{\'e}zard}, \citenamefont {Parisi},\ and\ \citenamefont
  {Zecchina}}]{Mezard2002}%
  \BibitemOpen
  \bibfield  {author} {\bibinfo {author} {\bibfnamefont {M.}~\bibnamefont
  {M{\'e}zard}}, \bibinfo {author} {\bibfnamefont {G.}~\bibnamefont {Parisi}},
  \ and\ \bibinfo {author} {\bibfnamefont {R.}~\bibnamefont {Zecchina}},\
  }\href {\doibase 10.1126/science.1073287} {\bibfield  {journal} {\bibinfo
  {journal} {Science}\ }\textbf {\bibinfo {volume} {297}},\ \bibinfo {pages}
  {812} (\bibinfo {year} {2002})}\BibitemShut {NoStop}%
\bibitem [{\citenamefont {M\'ezard}\ and\ \citenamefont
  {Zecchina}(2002)}]{Mezard2002b}%
  \BibitemOpen
  \bibfield  {author} {\bibinfo {author} {\bibfnamefont {M.}~\bibnamefont
  {M\'ezard}}\ and\ \bibinfo {author} {\bibfnamefont {R.}~\bibnamefont
  {Zecchina}},\ }\href {\doibase 10.1103/PhysRevE.66.056126} {\bibfield
  {journal} {\bibinfo  {journal} {Phys. Rev. E}\ }\textbf {\bibinfo {volume}
  {66}},\ \bibinfo {pages} {056126} (\bibinfo {year} {2002})}\BibitemShut
  {NoStop}%
\bibitem [{\citenamefont {Braunstein}\ \emph {et~al.}(2007)\citenamefont
  {Braunstein}, \citenamefont {M\'{e}zard},\ and\ \citenamefont
  {Zecchina}}]{Braunstein2007}%
  \BibitemOpen
  \bibfield  {author} {\bibinfo {author} {\bibfnamefont {A.}~\bibnamefont
  {Braunstein}}, \bibinfo {author} {\bibfnamefont {M.}~\bibnamefont
  {M\'{e}zard}}, \ and\ \bibinfo {author} {\bibfnamefont {R.}~\bibnamefont
  {Zecchina}},\ }\href {\doibase 10.1002/rsa.20057} {\bibfield  {journal}
  {\bibinfo  {journal} {Random Struct. Alg.}\ }\textbf {\bibinfo {volume}
  {27}},\ \bibinfo {pages} {201} (\bibinfo {year} {2007})}\BibitemShut
  {NoStop}%
\bibitem [{\citenamefont {Krz\c{a}ka{\l}a}\ \emph {et~al.}(2007)\citenamefont
  {Krz\c{a}ka{\l}a}, \citenamefont {Montanari}, \citenamefont
  {Ricci-Tersenghi}, \citenamefont {Semerjian},\ and\ \citenamefont
  {Zdeborov{\'a}}}]{Krzakala2007}%
  \BibitemOpen
  \bibfield  {author} {\bibinfo {author} {\bibfnamefont {F.}~\bibnamefont
  {Krz\c{a}ka{\l}a}}, \bibinfo {author} {\bibfnamefont {A.}~\bibnamefont
  {Montanari}}, \bibinfo {author} {\bibfnamefont {F.}~\bibnamefont
  {Ricci-Tersenghi}}, \bibinfo {author} {\bibfnamefont {G.}~\bibnamefont
  {Semerjian}}, \ and\ \bibinfo {author} {\bibfnamefont {L.}~\bibnamefont
  {Zdeborov{\'a}}},\ }\href {\doibase 10.1073/pnas.0703685104} {\bibfield
  {journal} {\bibinfo  {journal} {PNAS}\ }\textbf {\bibinfo {volume} {104}},\
  \bibinfo {pages} {10318} (\bibinfo {year} {2007})}\BibitemShut {NoStop}%
\bibitem [{\citenamefont {Fu}\ and\ \citenamefont {Anderson}(1986)}]{Fu1986}%
  \BibitemOpen
  \bibfield  {author} {\bibinfo {author} {\bibfnamefont {Y.}~\bibnamefont
  {Fu}}\ and\ \bibinfo {author} {\bibfnamefont {P.~W.}\ \bibnamefont
  {Anderson}},\ }\href {http://stacks.iop.org/0305-4470/19/i=9/a=033}
  {\bibfield  {journal} {\bibinfo  {journal} {J. Phys. A}\ }\textbf {\bibinfo
  {volume} {19}},\ \bibinfo {pages} {1605} (\bibinfo {year}
  {1986})}\BibitemShut {NoStop}%
\bibitem [{\citenamefont {Kanter}\ and\ \citenamefont
  {Sompolinsky}(1987)}]{Kanter1987}%
  \BibitemOpen
  \bibfield  {author} {\bibinfo {author} {\bibfnamefont {I.}~\bibnamefont
  {Kanter}}\ and\ \bibinfo {author} {\bibfnamefont {H.}~\bibnamefont
  {Sompolinsky}},\ }\href {http://stacks.iop.org/0305-4470/20/i=11/a=001}
  {\bibfield  {journal} {\bibinfo  {journal} {J. Phys. A}\ }\textbf {\bibinfo
  {volume} {20}},\ \bibinfo {pages} {L673} (\bibinfo {year}
  {1987})}\BibitemShut {NoStop}%
\bibitem [{\citenamefont {Zdeborov\'{a}}\ and\ \citenamefont
  {M\'{e}zard}(2006)}]{Zdeborova2006}%
  \BibitemOpen
  \bibfield  {author} {\bibinfo {author} {\bibfnamefont {L.}~\bibnamefont
  {Zdeborov\'{a}}}\ and\ \bibinfo {author} {\bibfnamefont {M.}~\bibnamefont
  {M\'{e}zard}},\ }\href {\doibase 10.1088/1742-5468/2006/05/P05003} {\bibfield
   {journal} {\bibinfo  {journal} {J. Stat. Mech. Theory Exp.}\ }\textbf
  {\bibinfo {volume} {2006}},\ \bibinfo {pages} {P05003} (\bibinfo {year}
  {2006})}\BibitemShut {NoStop}%
\bibitem [{\citenamefont {Fyodorov}\ and\ \citenamefont
  {Le~Doussal}(2014)}]{Fyodorov2014}%
  \BibitemOpen
  \bibfield  {author} {\bibinfo {author} {\bibfnamefont {Y.~V.}\ \bibnamefont
  {Fyodorov}}\ and\ \bibinfo {author} {\bibfnamefont {P.}~\bibnamefont
  {Le~Doussal}},\ }\href {\doibase 10.1007/s10955-013-0838-1} {\bibfield
  {journal} {\bibinfo  {journal} {J. Stat. Phys.}\ }\textbf {\bibinfo {volume}
  {154}},\ \bibinfo {pages} {466} (\bibinfo {year} {2014})}\BibitemShut
  {NoStop}%
\bibitem [{\citenamefont {M\'{e}zard}\ and\ \citenamefont
  {Parisi}(1987)}]{Mezard1987}%
  \BibitemOpen
  \bibfield  {author} {\bibinfo {author} {\bibfnamefont {M.}~\bibnamefont
  {M\'{e}zard}}\ and\ \bibinfo {author} {\bibfnamefont {G.}~\bibnamefont
  {Parisi}},\ }\href {\doibase 10.1051/jphys:019870048090145100} {\bibfield
  {journal} {\bibinfo  {journal} {J. Phys. (Paris)}\ }\textbf {\bibinfo
  {volume} {48}},\ \bibinfo {pages} {1451} (\bibinfo {year}
  {1987})}\BibitemShut {NoStop}%
\bibitem [{\citenamefont {Parisi}\ and\ \citenamefont
  {Rati\'{e}ville}(2002)}]{Parisi2002}%
  \BibitemOpen
  \bibfield  {author} {\bibinfo {author} {\bibfnamefont {G.}~\bibnamefont
  {Parisi}}\ and\ \bibinfo {author} {\bibfnamefont {M.}~\bibnamefont
  {Rati\'{e}ville}},\ }\href {\doibase 10.1140/epjb/e2002-00326-3} {\bibfield
  {journal} {\bibinfo  {journal} {Eur. Phys. J. B}\ }\textbf {\bibinfo {volume}
  {29}},\ \bibinfo {pages} {457} (\bibinfo {year} {2002})}\BibitemShut
  {NoStop}%
\bibitem [{\citenamefont {Caracciolo}\ \emph {et~al.}(2017)\citenamefont
  {Caracciolo}, \citenamefont {D'Achille}, \citenamefont {Malatesta},\ and\
  \citenamefont {Sicuro}}]{Caracciolo2017}%
  \BibitemOpen
  \bibfield  {author} {\bibinfo {author} {\bibfnamefont {S.}~\bibnamefont
  {Caracciolo}}, \bibinfo {author} {\bibfnamefont {M.~P.}\ \bibnamefont
  {D'Achille}}, \bibinfo {author} {\bibfnamefont {E.~M.}\ \bibnamefont
  {Malatesta}}, \ and\ \bibinfo {author} {\bibfnamefont {G.}~\bibnamefont
  {Sicuro}},\ }\href {\doibase 10.1103/PhysRevE.95.052129} {\bibfield
  {journal} {\bibinfo  {journal} {Phys. Rev. E}\ }\textbf {\bibinfo {volume}
  {95}},\ \bibinfo {pages} {052129} (\bibinfo {year} {2017})}\BibitemShut
  {NoStop}%
\bibitem [{\citenamefont {M\'{e}zard}\ and\ \citenamefont
  {Parisi}(2001)}]{Mezard2001}%
  \BibitemOpen
  \bibfield  {author} {\bibinfo {author} {\bibfnamefont {M.}~\bibnamefont
  {M\'{e}zard}}\ and\ \bibinfo {author} {\bibfnamefont {G.}~\bibnamefont
  {Parisi}},\ }\href {\doibase 10.1007/PL00011099} {\bibfield  {journal}
  {\bibinfo  {journal} {Eur. Phys. J. B}\ }\textbf {\bibinfo {volume} {20}},\
  \bibinfo {pages} {217} (\bibinfo {year} {2001})}\BibitemShut {NoStop}%
\bibitem [{\citenamefont {M\'{e}zard}\ and\ \citenamefont
  {Parisi}(2003)}]{Mezard2003}%
  \BibitemOpen
  \bibfield  {author} {\bibinfo {author} {\bibfnamefont {M.}~\bibnamefont
  {M\'{e}zard}}\ and\ \bibinfo {author} {\bibfnamefont {G.}~\bibnamefont
  {Parisi}},\ }\href {\doibase 10.1023/A:1022221005097} {\bibfield  {journal}
  {\bibinfo  {journal} {J. Stat. Phys.}\ }\textbf {\bibinfo {volume} {111}},\
  \bibinfo {pages} {1} (\bibinfo {year} {2003})}\BibitemShut {NoStop}%
\bibitem [{\citenamefont {M\'{e}zard}\ and\ \citenamefont
  {Montanari}(2009)}]{mezard2009information}%
  \BibitemOpen
  \bibfield  {author} {\bibinfo {author} {\bibfnamefont {M.}~\bibnamefont
  {M\'{e}zard}}\ and\ \bibinfo {author} {\bibfnamefont {A.}~\bibnamefont
  {Montanari}},\ }\href@noop {} {\emph {\bibinfo {title} {{Information,
  Physics, and Computation}}}},\ Oxford Graduate Texts\ (\bibinfo  {publisher}
  {OUP Oxford},\ \bibinfo {year} {2009})\BibitemShut {NoStop}%
\bibitem [{\citenamefont {Rivoire}(2005)}]{Rivoire2005}%
  \BibitemOpen
  \bibfield  {author} {\bibinfo {author} {\bibfnamefont {O.}~\bibnamefont
  {Rivoire}},\ }\href {http://stacks.iop.org/1742-5468/2005/i=07/a=P07004}
  {\bibfield  {journal} {\bibinfo  {journal} {J. Stat. Mech. Theory Exp.}\
  }\textbf {\bibinfo {volume} {2005}},\ \bibinfo {pages} {P07004} (\bibinfo
  {year} {2005})}\BibitemShut {NoStop}%
\bibitem [{\citenamefont {Crisanti}\ and\ \citenamefont
  {Leuzzi}(2014)}]{Crisanti2014}%
  \BibitemOpen
  \bibfield  {author} {\bibinfo {author} {\bibfnamefont {A.}~\bibnamefont
  {Crisanti}}\ and\ \bibinfo {author} {\bibfnamefont {L.}~\bibnamefont
  {Leuzzi}},\ }\enquote {\bibinfo {title} {Large deviations in disordered spin
  systems},}\ in\ \href@noop {} {\emph {\bibinfo {booktitle} {Large Deviations
  in Physics: The Legacy of the Law of Large Numbers}}},\ \bibinfo {editor}
  {edited by\ \bibinfo {editor} {\bibfnamefont {A.}~\bibnamefont {Vulpiani}},
  \bibinfo {editor} {\bibfnamefont {F.}~\bibnamefont {Cecconi}}, \bibinfo
  {editor} {\bibfnamefont {M.}~\bibnamefont {Cencini}}, \bibinfo {editor}
  {\bibfnamefont {A.}~\bibnamefont {Puglisi}}, \ and\ \bibinfo {editor}
  {\bibfnamefont {D.}~\bibnamefont {Vergni}}}\ (\bibinfo  {publisher} {Springer
  Berlin Heidelberg},\ \bibinfo {address} {Berlin, Heidelberg},\ \bibinfo
  {year} {2014})\ pp.\ \bibinfo {pages} {135--160}\BibitemShut {NoStop}%
\bibitem [{\citenamefont {Crisanti}\ \emph {et~al.}(1992)\citenamefont
  {Crisanti}, \citenamefont {Paladin}, \citenamefont {Sommers},\ and\
  \citenamefont {Vulpiani}}]{Crisanti1992b}%
  \BibitemOpen
  \bibfield  {author} {\bibinfo {author} {\bibfnamefont {A.}~\bibnamefont
  {Crisanti}}, \bibinfo {author} {\bibfnamefont {G.}~\bibnamefont {Paladin}},
  \bibinfo {author} {\bibfnamefont {H.-J.}\ \bibnamefont {Sommers}}, \ and\
  \bibinfo {author} {\bibfnamefont {A.}~\bibnamefont {Vulpiani}},\ }\href
  {\doibase 10.1051/jp1:1992213} {\bibfield  {journal} {\bibinfo  {journal}
  {{J. Phys. I France}}\ }\textbf {\bibinfo {volume} {2}},\ \bibinfo {pages}
  {1325} (\bibinfo {year} {1992})}\BibitemShut {NoStop}%
\bibitem [{\citenamefont {Parisi}\ and\ \citenamefont
  {Rizzo}(2008)}]{Parisi2008}%
  \BibitemOpen
  \bibfield  {author} {\bibinfo {author} {\bibfnamefont {G.}~\bibnamefont
  {Parisi}}\ and\ \bibinfo {author} {\bibfnamefont {T.}~\bibnamefont {Rizzo}},\
  }\href {\doibase 10.1103/PhysRevLett.101.117205} {\bibfield  {journal}
  {\bibinfo  {journal} {Phys. Rev. Lett.}\ }\textbf {\bibinfo {volume} {101}},\
  \bibinfo {pages} {117205} (\bibinfo {year} {2008})}\BibitemShut {NoStop}%
\bibitem [{\citenamefont {Parisi}\ and\ \citenamefont
  {Rizzo}(2009)}]{Parisi2009}%
  \BibitemOpen
  \bibfield  {author} {\bibinfo {author} {\bibfnamefont {G.}~\bibnamefont
  {Parisi}}\ and\ \bibinfo {author} {\bibfnamefont {T.}~\bibnamefont {Rizzo}},\
  }\href {\doibase 10.1103/PhysRevB.79.134205} {\bibfield  {journal} {\bibinfo
  {journal} {Phys. Rev. B}\ }\textbf {\bibinfo {volume} {79}},\ \bibinfo
  {pages} {134205} (\bibinfo {year} {2009})}\BibitemShut {NoStop}%
\bibitem [{\citenamefont {Parisi}\ and\ \citenamefont
  {Rizzo}(2010)}]{Parisi2010a}%
  \BibitemOpen
  \bibfield  {author} {\bibinfo {author} {\bibfnamefont {G.}~\bibnamefont
  {Parisi}}\ and\ \bibinfo {author} {\bibfnamefont {T.}~\bibnamefont {Rizzo}},\
  }\href {\doibase 10.1103/PhysRevB.81.094201} {\bibfield  {journal} {\bibinfo
  {journal} {Phys. Rev. B}\ }\textbf {\bibinfo {volume} {81}},\ \bibinfo
  {pages} {094201} (\bibinfo {year} {2010})}\BibitemShut {NoStop}%
\bibitem [{\citenamefont {Kondor}(1983)}]{Kondor1983}%
  \BibitemOpen
  \bibfield  {author} {\bibinfo {author} {\bibfnamefont {I.}~\bibnamefont
  {Kondor}},\ }\href {http://stacks.iop.org/0305-4470/16/i=4/a=006} {\bibfield
  {journal} {\bibinfo  {journal} {J. Phys. A}\ }\textbf {\bibinfo {volume}
  {16}},\ \bibinfo {pages} {L127} (\bibinfo {year} {1983})}\BibitemShut
  {NoStop}%
\bibitem [{\citenamefont {W{\"a}stlund}(2010)}]{Wastlund2010}%
  \BibitemOpen
  \bibfield  {author} {\bibinfo {author} {\bibfnamefont {J.}~\bibnamefont
  {W{\"a}stlund}},\ }\href@noop {} {\bibfield  {journal} {\bibinfo  {journal}
  {Acta Mathematica}\ }\textbf {\bibinfo {volume} {204}},\ \bibinfo {pages}
  {91} (\bibinfo {year} {2010})}\BibitemShut {NoStop}%
\bibitem [{\citenamefont {M\'{e}zard}\ and\ \citenamefont
  {Parisi}(1986{\natexlab{b}})}]{Mezard1986a}%
  \BibitemOpen
  \bibfield  {author} {\bibinfo {author} {\bibfnamefont {M.}~\bibnamefont
  {M\'{e}zard}}\ and\ \bibinfo {author} {\bibfnamefont {G.}~\bibnamefont
  {Parisi}},\ }\href {http://iopscience.iop.org/0295-5075/2/12/005} {\bibfield
  {journal} {\bibinfo  {journal} {EPL}\ }\textbf {\bibinfo {volume} {2}},\
  \bibinfo {pages} {913} (\bibinfo {year} {1986}{\natexlab{b}})}\BibitemShut
  {NoStop}%
\bibitem [{\citenamefont {Parisi}\ and\ \citenamefont
  {Rati\'{e}ville}(2001)}]{Parisi2001}%
  \BibitemOpen
  \bibfield  {author} {\bibinfo {author} {\bibfnamefont {G.}~\bibnamefont
  {Parisi}}\ and\ \bibinfo {author} {\bibfnamefont {M.}~\bibnamefont
  {Rati\'{e}ville}},\ }\href {\doibase 10.1007/PL00011144} {\bibfield
  {journal} {\bibinfo  {journal} {Eur. Phys. J. B}\ }\textbf {\bibinfo {volume}
  {22}},\ \bibinfo {pages} {229} (\bibinfo {year} {2001})}\BibitemShut
  {NoStop}%
\bibitem [{\citenamefont {Charalambides}(2002)}]{Charalambides2002}%
  \BibitemOpen
  \bibfield  {author} {\bibinfo {author} {\bibfnamefont {C.}~\bibnamefont
  {Charalambides}},\ }\href@noop {} {\emph {\bibinfo {title} {Enumerative
  Combinatorics}}},\ Discrete Mathematics and Its Applications\ (\bibinfo
  {publisher} {Taylor \& Francis},\ \bibinfo {year} {2002})\BibitemShut
  {NoStop}%
\end{thebibliography}%
\end{document}